\documentclass{aa}  

\usepackage{graphicx}
\usepackage{txfonts}
\begin{document}

   \title{ Detailed abundances in a sample of very metal poor stars }

   \subtitle{}

   \author{P. Fran\c cois
          \inst{1,2}
          \and
          S. Wanajo \inst{3,4, 5}
           \and
           E. Caffau \inst{6}
           \and
           N. Prantzos \inst{7}
           \and 
           W. Aoki \inst{9}
           \and
           M. Aoki \inst{8}
            \and
           P. Bonifacio \inst{6}  
           \and 
           M. Spite \inst{6}
           \and 
           F. Spite \inst{6}
           \  \thanks{Based on data collected at Subaru Telescope, which is operated by the National Astronomical Observatory of Japan.}  
          }

   \institute{GEPI, Observatoire de Paris, PSL Research University, CNRS, 61 Avenue de l'Observatoire, 75014 Paris, France \\
              \email{patrick.francois@obspm.fr}
         \and
             UPJV, Universit\'e de Picardie Jules Verne, 33 rue St Leu, 80080 Amiens, France 
          \and
          Max Planck Institute for Gravitational Physics (Albert Einstein Institute), Am M\"uhlenberg 1, Potsdam-Golm, 14476, Germany
	\and
          iTHEMS Research Group, RIKEN, Wako, Saitama 351-0198, Japan
          \and
          Department of Engineering and Applied Sciences, Faculty of Science and Technology, Sophia University, 7-1 Kioicho, Chiyoda-ku, Tokyo 102-8554, Japan
          \and
          GEPI, Observatoire de Paris, PSL Research University, CNRS, Place Jules Janssen, 92190 Meudon, France    
           \and
          Institut d’Astrophysique de Paris, UMR7095 CNRS, Univ. P. \& M. Curie, 98bis Bd. Arago, F-75104 Paris, France
          \and
            European Southern Observatory, Karl-Schwarzschild-Str. 2   85748 Garching bei Muenchen, Germany
          \and
          National Observatory of Japan, Mitaka, Tokyo, Japan
               }

   \date{Received 26 Mar 2020; accepted  8 Jul 2020}

 
  \abstract
   { Unevolved metal poor stars are the witness of the early evolution of the Galaxy.  The determination of their detailed chemical composition  is an important tool to understand the chemical history of our Galaxy. The study of their chemical composition can also be used to constrain the  nucleosynthesis of the first generation of supernovae that enriched the interstellar medium. 
  }
   { The aim is to observe a sample of  extremely metal poor stars (EMP stars)  candidates selected from SDSS DR12 release and  determine their chemical composition.}
   { We obtained high resolution spectra of  a sample of five stars using   HDS  on Subaru telescope and used standard 1D models to compute the  abundances. The stars we analysed have a metallicity [Fe/H]  between -3.50 dex and -4.25 dex . }
   {We confirm that the five metal poor candidates selected from low resolution spectra are very metal poor. 
  We present, the discovery of a new ultra  metal-poor star (UMP star) with a metallicity  of  [Fe/H]= -4.25 dex (SDSS~J1050032.34$-$241009.7). We measured in this star an upper limit of 
  lithium ( log(Li/H) $\le$ 2.0.  We found that the 4 most metal poor stars of our sample have a lower lithium abundance than the Spite plateau lithium value.  We obtain upper limits for carbon in the sample of stars. None of them belong to the high carbon band. We  measured abundances of Mg and Ca  in most of the stars and found three new   $\alpha$-poor stars. }
   {}


   \maketitle
%

\section{Introduction}

Metal poor stars are the witnesses of the early evolution of our Galaxy. 
They provide important clues to the formation of the first objects in the Universe.  The detailed abundance analysis of their atmosphere reveal observational 
details that can be compared to theoretical models for nucleosynthesis in the first metal poor massive stars exploding as supernovae.
The trends of their abundance ratios as a function of the metallicity can be used to trace the chemical history of our Galaxy. 
 A thorough discussion of the main scientific goals of the study of metal poor stars 
can be found in several review articles \citep[see for example][]{beers2005, frebel2015}.

First misidentified as early type stars \citep[e.g.][]{1935ApJ81_187A}, the metal poor stars revealed their true
nature as old metal-poor stars only through high resolution spectroscopic analysis
\citep{1951ApJ114_52C}. The catalogue of high velocity stars of Roman(1955) was extensively use as source of low metallicity stars 
\citep[e.g.][]{1957ApJ...126..281G,1959ApJ...129..720W,1962ApJS....6..407W,1963ApJ...137..280W}.
Although this field of research was largely dominated by the work of astronomers in the United States, a growing interest began also in Europe \citep[e.g][]{1963ZA.....56..207B,cayrel1964}.
 
The search for the most metal poor stars has made great progresses with the survey work of \citet{beers1985} and the discovery of several very metal poor stars with [Fe/H] $\le$ -3.00 dex \citep{norris1993}.  
Their survey was making use of a temperature index  based on the  H $\beta$  hydrogen absorption line and a metallicity index  from the strong Ca H\&K lines  which can be detected and measured on low resolution spectra
even at low metallicity.  The high resolution follow-up observations of metal poor candidates led to a large number of publications among them the series of articles from the First Stars ESO Large programme (Prog ID 165.N-0276) published 
by the Cayrel's group \citep[and references therein]{bonifacio2009, cayrel2004}.  
They also  showed how rare are the very metal poor stars as revealed from the metallicity distribution function,  implying larger and and deeper survey to find new very metal poor candidates. 
The Hamburg ESO survey whose first aim was the search for new quasars has also been very successful in the search  and the discovery  of several stars with [Fe/H] with metallicity lower than -4.00 dex  \citep[and references therein]{christlieb2008}. 
More recently,  the Sloan Digital Sky Survey (SDSS) spectroscopic survey has been used to identify new metal poor candidates \citep{helmi2003}.   \citet{ludwig2008}  have developed an analysis tool that allows us to estimate the metallicity of Turn Off (TO) stars from the low resolution SDSS spectra. Follow up observations on large telescope has been used to confirm their metallicity and measure abundance ratios. This detection method has been quite successful and led to the discovery of several 
interesting very metal poor stars  \citep{caffau2011a, caffau2011b, caffau2012, caffau2014, bonifacio2015, caffau2016, bonifacio2018, francois2018}. 
Other surveys have also been used to detect metal poor stars, as the Apache Point Observatory Galactic Evolution Experiment (APOGEE,  \citet{ majewski2016}), a  survey toward the galactic centre, where the stellar density allows large multiplex spectroscopic observations. 
We could also mention the Radial Velocity Experiment (RAVE) survey which  first aim  is  the study of the galactic dynamics from a radial velocity census of stars \citep{steinmetz2006}, the Skymapper Sky Survey  \citep{keller07, wolf2018}  performing wide field imaging in five wide bands and a narrow band centred on Ca H\&K absorption lines  using the Skymapper telescope. The photometric SkyMapper Southern Sky Survey \citep{keller07} discovered two
  of the most extremely iron-poor stars: SMSS\,J031300.36-670839.3 \citep{keller14}
  and SMSS\,J160540.18-144323.1 \citep{nordlander19}.
  More recently,  the  PRISTINE survey \citep{starkenburg2017} based on Canada France Hawaii telescope (CFHT) large field imaging uses a dedicated narrow band filter centred  on Ca H\&K absorption lines,  combined with SDSS broad-band {\it g} and {\it i}  photometry.

The common point between these different sources of metal poor stars candidates, is that they require high resolution spectroscopic follow-up observations to confirm their low metallicity and determine their detailed chemical composition. 

 From the recent analysis of  SDSS DR12 data, we have detected new extremely metal poor  candidates that have never been observed at high resolution. In this article, we report  the detailed analysis of five new extremely  metal poor candidates observed with the  High Dispersion Spectrograph (HDS) installed on the Subaru telescope   atop Mauna Kea volcano  in Hawaii.  Similar observations have been conducted for a  second set of stars  visible from  the southern hemisphere using   the X-SHOOTER spectrograph installed on the UT2 (KUEYEN)  at the ESO very large telescope (VLT) on Cerro  Paranal in Chile  in the framework of a french-japanese collaboration.  The results have been published in  \citet{francois2018}.

\section{Observation }

The observations have been carried out  with HDS installed on the Subaru telescope \citep{noguchi2002}. The wavelength coverage goes from  4084 \AA ~to  6892 \AA  . 
A binning 2x2 has been adopted leading to a resolution of about 40000.  The logbook of the observations is given in table \ref{obslog} . Standard data reduction procedures were carried out with the IRAF Echelle package \footnote{ IRAF is distributed by the National Optical Astronomy Observatories, which is operated by the Association of Universities for Research in Astronomy, Inc. under cooperative agreement with the National Science Foundation}. Care has been taken to remove the sky background as most of the exposures were affected by the moon illumination.

\begin{table*}
 \caption{Observation log } 
\label{obslog}
\centering
\begin{tabular}{l c c c c }
\hline\hline
  Object          &  g magnitude  &  Observation date &  Exp. time  [s]  &S/N @ 480 nm  \\      
\hline

SDSS~J081554.26$+$472947.5  &  17.06  &     2017-04-05T06:03  & 3600 & 25 \\        
SDSS~J081554.26$+$472947.5  &   17.06 &     2017-04-05T07:04  & 3600 & 30 \\        
SDSS~J081554.26$+$472947.5  &   17.06 &     2017-04-05T08:05  & 3600 & 30 \\        
SDSS~J091753.19$+$523004.9 &  18.58  &     2017-04-06T05:55 & 3600&  15 \\           
SDSS~J091753.19$+$523004.9   &  18.58 &     2017-04-06T06:56 & 3600&   15 \\           
SDSS~J091753.19$+$523004.9   &  18.58 &     2017-04-06T07:57 & 3600&   15 \\            
SDSS~J105002.34$+$242109.7    &18.04  & 2017-04-05T09:10  & 3600 &  20\\                
SDSS~J105002.34$+$242109.7    &18.04  & 2017-04-05T10:11  & 3600 & 20 \\              
SDSS~J105002.34$+$242109.7    &18.04  &     2017-04-06T09:05 & 3600& 25 \\             
SDSS~J105002.34$+$242109.7    & 18.04  &    2017-04-06T10:05 & 3600&  20 \\             
SDSS~J105002.34$+$242109.7    & 18.04  &    2017-04-07T05:46 & 3600 & 25 \\              
SDSS~J105002.34$+$242109.7     & 18.04  &   2017-04-07T06:47 & 3335 & 25 \\              
SDSS~J124304.19$-$081230.6     &  18.25 &  2017-04-05T11:22  & 3600 & 15 \\            
SDSS~J124304.19$-$081230.6     &  18.25 &  2017-04-05T12:23  & 3600 & 10 \\           
SDSS~J124304.19$-$081230.6     &  18.25 &    2017-04-06T11:11 & 3600&  20 \\              
SDSS~J124304.19$-$081230.6     & 18.25  &    2017-04-06T12:12 & 3600&  15 \\              
SDSS~J124304.19$-$081230.6      & 18.25  &   2017-04-07T08:00 & 3600 & 20 \\              
SDSS~J124304.19$-$081230.6      &  18.25 &   2017-04-07T09:00 & 3600 & 20 \\              
SDSS~J153346.28$+$155701.8    & 16.91   &   2017-04-05T13:29 & 3000 & 20 \\           
SDSS~J153346.28$+$155701.8    & 16.91   &   2017-04-05T14:20 & 3000 & 20\\            
SDSS~J153346.28$+$155701.8    & 16.91   &   2017-04-06T13:16 & 3300 & 20 \\              
SDSS~J153346.28$+$155701.8    & 16.91  &    2017-04-06T14:11 & 3300 & 25 \\              

  \hline
  \end{tabular}
   \end{table*}

\begin{figure*}
   \centering
  \includegraphics[width=12cm]{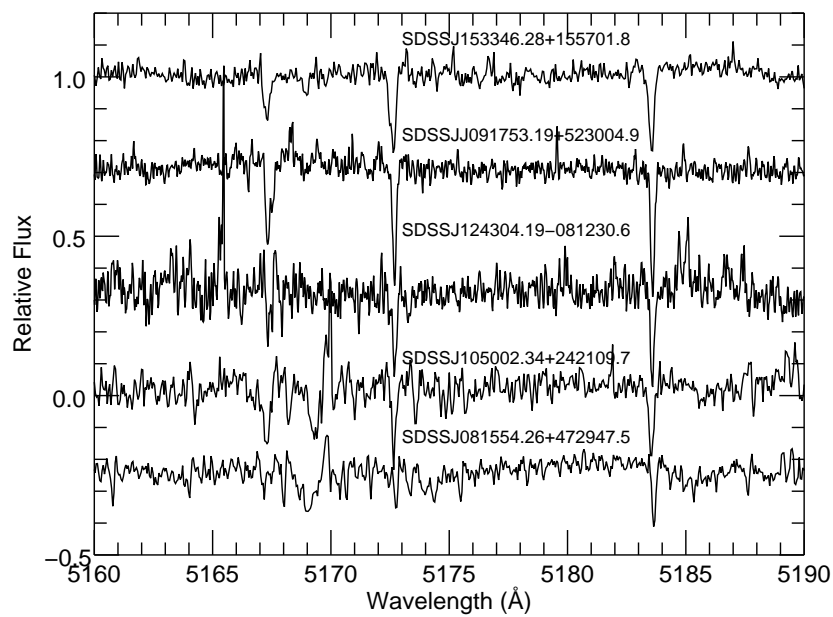}
      \caption{Parts of the HDS Subaru spectra centred on the magnesium triplet. The continuum of the four spectra located in the lower part of the plot has
      been shifted downwards for clarity. The two lower spectra reveal the presence of residuals of the sky spectrum in the region 5168-5170 $\AA$ and 5185-5186 $\AA$ }
              
         \label{Fig_spectra}
   \end{figure*}

Fig. \ref{Fig_spectra}  shows the spectra of the stars of our sample centred on the magnesium triplet. 
The continuum level of  the four spectra located in the lower part of the plot has been shifted for clarity. The spectra are presented with decreasing metallicity from the top to the bottom of the figure.

\section{Stellar parameters}

  \begin{figure*}
   \centering
   \includegraphics[width=10cm]{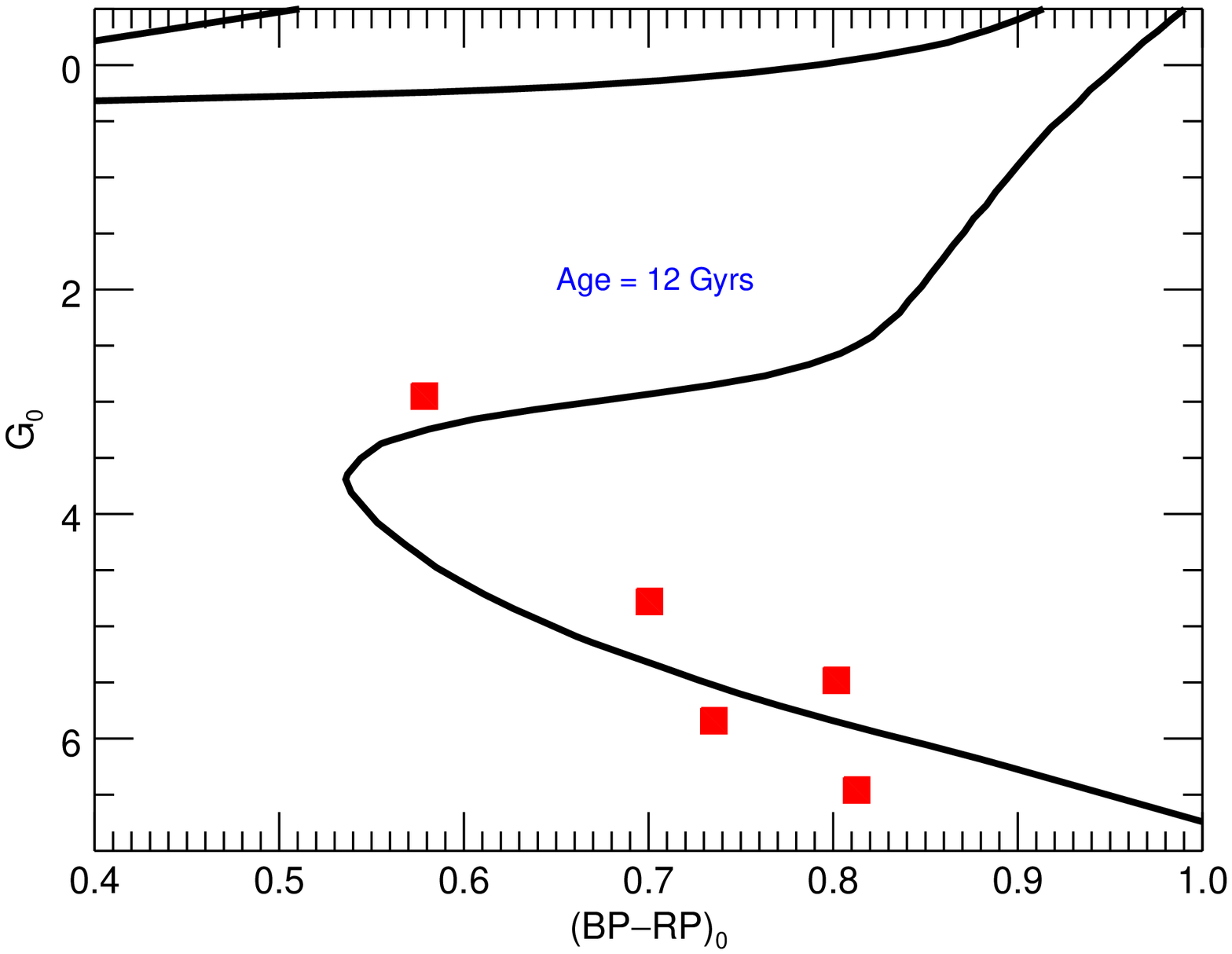}
      \caption{ Comparison of the location of the stars and a PARSEC \citep{parsec} isochrone
of metallicity --2.5 and age 12 Gyr on a $G_{0}$  magnitude versus $(BP-RP)_{0}$ magnitude diagram. The  isochrone on the plot has been computed for an age of 12 Gyr and a metallicity of -2.5 dex.   BP-RP (BP and RP are the magnitudes measured respectively by the two  low resolution spectrographs, the Blue Photometer (BP) and the Red Photometer (RP)  onboard the Gaia satellite) versus the absolute G magnitude.  
 The red symbols represent our stars. 
 }             
         \label{gaia_cmd}
   \end{figure*}

The stellar parameters have been derived  taking into account the SDSS photometry.

The effective temperatures in Table \ref{stellar_parameters} have  been computed by \citet{caffau2013} . 
The effective temperature has been derived from the photometry, using the 
$(g-z)_{0}$ colour and the calibration described in \citet{ludwig2008} taking into account  the 
reddening according to the \citet{schlegel1998} extinction maps and corrected
as in \citet{bonifacio2000}. The Gaia parallaxes  \citep[Gaia Collaboration et al. 2018]{arenou2018} for our stars are  imprecise.
 SDSSJ081554.26$+$472947.5  and SDSSJ091753.19$+$523004.9~  parallaxes have a relative error  smaller than $\sim 20$\%.
 SDSSJ124304.19$-$081230.6~  and  SDSSJ153346.28$+$155701.8~  parallaxes have a relative error  of the order of  $\sim 50$\%.
One star  (SDSSJ105002.34$+$242109.7 ) has a negative parallax. 
In order to get some insight in the luminosity of this star, we used the distance estimates of \citet{BJ18}.
On Figure \ref{gaia_cmd}, we plotted  $(BP-RP)_{0}$  versus the absolute $G_{0}$ magnitude (BP and RP are the magnitudes measured respectively by the two  low resolution spectrographs, the Blue Photometer (BP) and the Red Photometer (RP)  onboard the Gaia satellite).  
The red symbols represent our stars assuming $$E(PB - RP) = (BP-RP) - 1.289445 \times E_{(B-V)\_SandF}$$  and $$G_{0} =  G - 0.85926 \times AV_{SandF}$$ where $E_{(B-V)\_{SandF}}$    and $AV_{SandF}$ are obtained from the maps of \citet{schlafly2011}.  We have also plotted on Figure \ref{gaia_cmd}  a PARSEC (PAdova and TRieste Stellar Evolution Code)  \citep{parsec} isochrone of metallicity -2.5 dex and an age of 12 Gyrs. The location of the stars clearly reveals that the stars are not giant stars. Indeed, the $G_{0}$  for our stars are between 3.0 and 6.5 which correspond to log~g ranging from $\simeq$ 3.5 to 4.7 based on the isochrone. 
It confirms that four of our stars are dwarf stars and one seems slightly evolved. Adopting log~g = 4.00 for the gravity of our stars is  suited for our analysis. 
 We remind that the selection of the stars is based on the dereddened $(g - z)$ and $(u - g)$ colours: $0.18 \le (g - z)_{0} \leq 0.70$ and $(u - g)_{0} \geq 0.70$. As discussed in \citet{bonifacio2012}, this selects the stars of the halo turn-off and excludes the majority of the white dwarf stars. A microturbulent velocity fo 1.5 km/s suitable for stars with log g =4 dex  has been adopted  following the results of \citet{barklem2005}.  The metallicities shown in Table \ref{stellar_parameters} 	have been computed by \citet{caffau2013}  using the code MyGisFoS \citep{sbordone2010} and the SDSS spectra of the stars.

 \begin{table*}
 \caption{Adopted stellar parameters for the list of targets. The last column gives the measured radial velocity of the stars after correction from the barycentric velocity.} 
\label{stellar_parameters}
\centering
\begin{tabular}{l c c c c c}
\hline\hline
  Object          &  ${\rm T}_{eff}$ & log~g  & [Fe/H]   & $\xi$ (km/s)  &  $v_{r}$ (km/s)\\      
\hline

SDSS~J081554.26$+$472947.5  &  6066   &  4.00   &   -4.00 & 1.5    &    -120 \\    
SDSS~J091753.19$+$523004.9   & 5858   &  4.00 &   -3.50 & 1.5       &   -37 \\    
SDSS~J105002.34$+$242109.7   & 5682   &  4.00    &   -4.00 & 1.5     &   -126\\    
SDSS~J124304.19$-$081230.6    & 5488   &  4.00    &   -3.50 & 1.5    &    154 \\    
SDSS~J153346.28$+$155701.8   & 6375   &  4.00   &   -3.50  & 1.5     &   -283 \\    

  \hline
  \end{tabular}
   \end{table*}

              

  \begin{figure*}
   \centering
  \includegraphics[width=12cm]{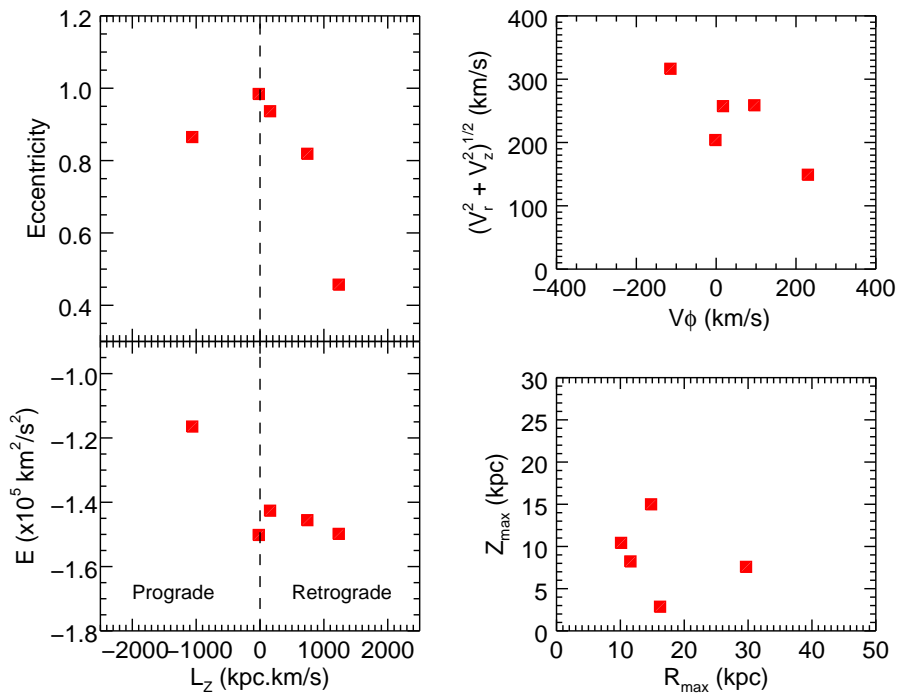}
      \caption{The orbital parameters for the stars.The eccentricity is computed as  e= $(R_{max}  - R_{min})/(R_{max}  + R_{min})$. }
              
         \label{OrbitEeccent}
   \end{figure*}

\begin{table*}
 \caption{Orbit parameters.} 
\label{orbit_parameters}
\centering
\begin{tabular}{l c c c c c c c c }
\hline\hline
  Object                                         &    $R_{min}$  & $R_{max}$ &   $Z_{max}$ &    Energy & $L_{Z}$ &  X  &     Y  &       Z \\
  \hline
SDSS~J081554.26$+$472947.5  &  0.08  &   10.11 & 10.42  &  -150171 &   -19.28 &10.06 & -0.26 &1.26  \\
SDSS~J091753.19$+$523004.9   & 0.49  &  14.80  & 14.99  &-142622 &156.89  & 9.21 &-0.27 &0.99 \\
SDSS~J105002.34$+$242109.7   &  2.15 &  29.68 & 7.58 &   -116458& -1061.98 &9.25 &0.65 &  2.38 \\
SDSS~J124304.19$-$081230.6    & 1.61  & 16.21  & 2.86 &-145572 &740.35 & 7.69  &0.92 & 1.50   \\
SDSS~J153346.28$+$155701.8   &  4.31 &  11.57  & 8.23 &-149829 &1232.96 & 5.18 &-1.40 &  4.10 \\
\hline
  \end{tabular}
\begin{tabular}{l c c c c c c c c c }
  Object                             &$V_{X}$   &    $V_{Y}$   &   $ V_{Z} $    &  $R$     &   $\phi$  &   $V_{r}$   &    $V_{Z}$  &   $V_{\phi}$\\
  \hline
SDSS~J081554.26$+$472947.5  &  18.53 &  -2.40 & -203.03 &10.07 &  358.53 & 18.59 &   -203.03 & -1.91  \\ 
SDSS~J091753.19$+$523004.9   &   -171.78 & 22.03 &  191.17 &  9.21 &  358.33 & -172.35&  191.17&  17.03     \\ 
SDSS~J105002.34$+$242109.7   &   -305.73& -136.25& -35.29 & 9.27 &  4.01 & -314.50 & -35.29 & -114.55    \\ 
SDSS~J124304.19$-$081230.6    &   239.80 & 124.96 &  54.68 &  7.74 &     6.82 & 252.94 &  54.68 &  95.59   \\ 
SDSS~J153346.28$+$155701.8   &    192.12 &  186.05 &  -59.38 & 5.36 &   344.83& 136.74&  -59.39 & 229.84  \\

  \hline
  \end{tabular}
   \end{table*}

From the radial velocities given in Table \ref{stellar_parameters} and using the publicly licensed code \rm{GalPot}  \footnote{
GalPot can be found at  https://github.com/PaulMcMillan-Astro/GalPot} , which is described by \citet{dehnen1998}, we computed the kinematic properties of our sample of stars we have analysed and some of the quantities derived from their Galactic orbits. The results are shown in Table \ref{orbit_parameters}.   In this table,  $L_{z}$ is the angular momentum, R is the galactocentric radius (cylindrical),
      $R_{min}$ and $R_{max}$ are respectively the minimum and the maximum values of the galactocentric radius (cylindrical) of the orbit, 
 E is the Energy of the orbit.  $Z_{max}$ is the maximum galactocentric height of the orbit. 
The space velocities (U, V, W) are with respect to the Local Standard of Rest, U is positive towards the Galactic anti-centre,V in the direction of the Galactic rotation and W is perpendicular to the Galactic plane, positive in the northern Galactic hemisphere. We list also the mean specific angular momentum (angular momentum per unit mass) for the stars along their orbits, in units of kpc $\times$ km s$^{-1}$. 
 In Fig. \ref{OrbitEeccent} the Toomree diagram and  the orbital characteristics of our sample are presented. 4 of the stars have large eccentricities and high values of $Z_{max}$ indicating that they are halo stars. 
The  remaining star with  the low $Z_{max}$=2.81 Kpc has an eccentricity of 0.82 and also likely a halo star.

\section{Analysis}

We carried out a classical 1D LTE analysis using OSMARCS model  atmospheres  \citep{gustafsson1975,   gustafsson2003,   gustafsson2008,  plez1992, edvardsson1993}. The abundances  used  in  the  model  atmospheres  were  solar-scaled with respect to the  \citet{grevesse2000} solar abundances, except for the $\alpha$-elements that are enhanced by 0.4 dex. We corrected the resulting abundances by taking into account the difference  between  \citep{grevesse2000} and  Caffau  et  al.(2011b), Lodders et al. (2009) solar abundances.
 To summarise, the solar abundances adopted for this work are log(C/H)$_{\odot}$=8.50,     log(Mg/H)$_{\odot}$=7.54,   log(Ca/H)$_{\odot}$=6.33, log(Fe/H)$_{\odot}$=7.52, log(Sr/H)$_{\odot}$=2.92 and log(Ba/H)$_{\odot}$=2.17. 

The abundance analysis was performed using the LTE spectral  line  analysis  code  turbospectrum  \citep{alvarez1998,    plez2012}, which treats scattering in detail. The carbon abundance was determined by fitting the CH band near to 430 nm (G band). The molecular data that correspond to the CH band are described in \citet{hill2002}.
The abundances have been determined by matching a synthetic spectrum centred on each line of interest to the observed spectrum.  Table \ref{linelist}  gathers the list of lines which have been used to measure the abundances or evaluate upper limits in our sample of stars. 

\begin{table*}
 \caption{List of  absorption lines used to determine the abundances} 
\label{linelist}
\centering
\begin{tabular}{l r r r }
\hline\hline
 Element &  Wavelength ($\AA$)   & \rm{$\chi_{esc}$} & log~gf \\
 
 \hline
 LiI            &6707.761   &   0.00  &  -0.009  \\
 LiI            &6707.912   &  0.00   &  -0.309  \\
 CH band &    4315      &             &              \\
 CH band &   4324       &             &             \\
 MgI          &5172.698   &2.71   &  -0.38      \\
 MgI          &5183.619    &2.72   &  -0.16      \\
CaI            & 4226.740   & 0.00    &  +0.24    \\   
FeI     &  4202.040 & 1.48   & -0.70  \\
FeI    &   4260.486  & 2.40   & -0.02  \\
FeI    &  4271.164  & 2.45   & -0.35   \\
FeI    &  4325.775  & 1.61   & -0.01    \\
FeI    &  4383.557    &1.48     &  0.20       \\
FeI    &  4404.761  & 1.56   & -0.14   \\
FeI    &  4415.135  & 1.61   & -0.61  \\ 
FeI     & 5269.550   & 0.86    &  -1.32      \\
SrII            & 4215.520   & 0.00    &  -0.17     \\  
BaII           & 4554.036   & 0.00   &   +0.16     \\
   \hline
  \end{tabular}
   \end{table*}

\section{Errors}

\begin{table*}
 \caption{Estimated errors in the element abundance ratios [X/Fe] for the star  SDSS~J091934.08$+$524014.0. The other stars give similar results } 
\label{errors}
\centering
\begin{tabular}{l c c c }
\hline\hline
  [X/Fe]            &   $\Delta T_{eff}$ =  100~K  & $\Delta$ log~g = 0.5~dex   &$\Delta$  $ v_{t}$ =   0.5~km/s \\     
\hline
C    &   0.2  &   0.2  &     0.1    \\
Mg &   0.1  &  0.15 &  0.15  \\
Ca &   0.1  &  0.1  &  0.15 \\  
Sr &   0.1  &  0.2  &  0.25  \\
Ba &   0.1  &  0.2   &  0.3  \\
  \hline
  \end{tabular}
   \end{table*}

Table \ref{errors} lists the computed errors in the elemental abundances ratios due to typical uncertainties in the stellar parameters. The errors were estimated varying  $T_{eff}$ by $\pm$ 100~K, log~g  by $\pm$  0.5~dex and $ v_{t}$  by $\pm$ 0.5 dex in the model atmosphere of  SDSS~J091934.08$+$524014.0, other stars give similar results. In this star, we could measure the Mg,  Ca  and set limits for Li, Sr, and Ba  abundances.  The main uncertainty comes from the error in the placement of the continuum when the synthetic line profiles are matched to the observed spectra. In particular, residuals from the  sky subtraction may lead to a decrease of the S/N ratio. As the final spectra are build from the the combination of several exposure taken at different epochs hence different barycentric velocities, the features sky residuals are smoothed and degrade the S/N of the spectra. 
This error is of the order of 0.1 to 0.2  depending on the S/N ratio of the spectrum and   the species under consideration, the largest value being for the neutron capture elements.  When several lines are available, the typical line to line scatter for a given elements is 0.1 to 0.2 dex.

\section{Results and discussion}

 \begin{table*}
 \caption{Lithium and  carbon abundances. $\alpha$ and  neutron-capture element  abundance ratios for the first 4 rows. The [Fe/H] abundance for  SDSS~J081554.26$+$472947.5  
 is an upper limit. Therefore, the abundances are given  as A(X) for Li and C  and  [X/H] for Fe, Mg, Ca, Sr and Ba.}   
\label{ratios}
\centering
\begin{tabular}{l r r r r r r r  }
\hline\hline
    
            Object                                 &  [Fe/H]  &          A(Li)       &  A(C)       & [Mg/Fe] &  [Ca/Fe]      &   [Sr/Fe]         & [Ba/Fe]        \\    
 \hline 
SDSS~JJ091753.19$+$523004.9   & -3.70             & $\le$ 1.8 &  $\le$ 6.7 &  0.46   &             0.27 &  $\le$  -0.42   & $\le$  ~0.23   \\    
SDSS~J105002.34$+$242109.7     &- 4.25             & $\le$ 2.0 &  $\le$ 6.8 &  0.41   & $\le$   -0.08 &  $\le$  ~0.53  & $\le$  ~1.18   \\    
SDSS~J124304.19$-$081230.6     &- 4.05              & $\le$ 1.7 &  $\le$ 7.0 &  0.21   &            -0.28 &  $\le$  -0.67   & $\le$  ~0.08   \\    
SDSS~J153346.28$+$155701.8    & -3.50              & $\le$ 2.2 &  $\le$ 7.0 &   0.26  &             0.07 &  $\le$  -0.22   &  $\le$  -0.17   \\    
 \hline\hline
            Object                                 &  [Fe/H]  &          A(Li)       &  A(C)       & [Mg/H] &  [Ca/H]      &   [Sr/H]         &  [Ba/H]        \\    
\hline            
SDSS~J081554.26$+$472947.5    &  $\le$  -4.10   & $\le$ 2.0 & $\le$  7.3 &  -3.67  & $\le$   -4.13 &  $\le$ ~-3.42   & $\le$   -2.77   \\    
\hline            
 
  \end{tabular}
   \end{table*}

The abundance and upper limit  results  for the sample of stars of this programme  have been gathered in Tables \ref{ratios}. For lithium and carbon, the log(X/H) is given  whereas [X/Fe] results are presented for magnesium, calcium, strontium and barium. For the star SDSS~J081554.26$+$472947.5, the abundance ratios are given assuming [Fe/H] {\bf <} -4.10 dex.  This upper limit of the [Fe/H] has been derived using the strongest FeI line available in our spectrum.

\subsection {Lithium}

\begin{figure*}
   \centering
  \includegraphics[width=12cm]{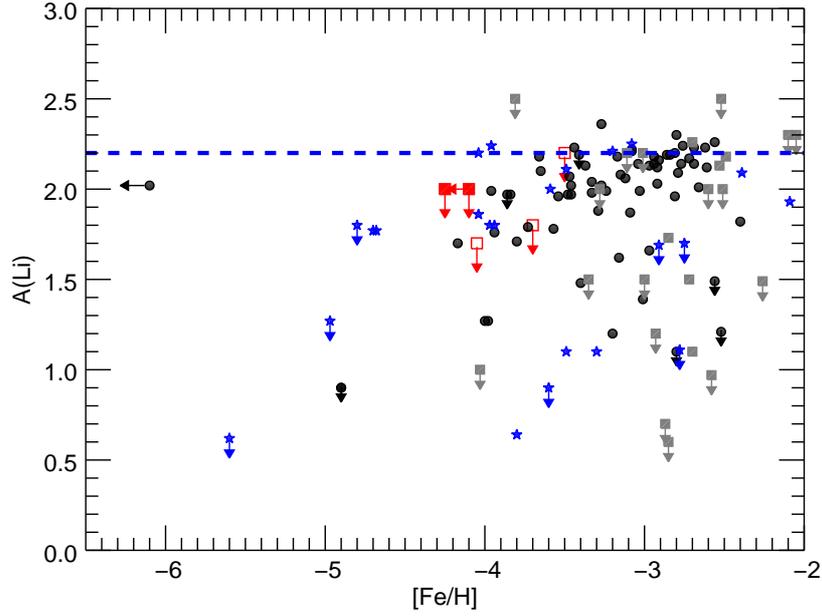}
      \caption{Lithium abundance in unevolved EMP stars.  The filled circles refer to carbon normal stars. CEMP stars of the low- and high-carbon bands are shown  as blue stars and grey rectangles.  Upper limits of the programme stars are shown in red.  Open red symbols represent the  stars for which we found with low [n-capture/Fe] upper limit abundances. 
      The blue dashed line represent the Spite Plateau as determined
 by \citet{sbordone2010}.  Details about literature data (black and grey symbols) can be found in \citet{bonifacio2018}. The star SDSSJ002314.00$+$030758.07 with [Fe/H] $<$ -6.10 dex  from \citet{aguado2019} has been added.
              }
         \label{Fig_Lithium}
   \end{figure*}

In Fig. \ref{Fig_Lithium} we plotted the abundance of lithium as a function of [Fe/H]. Our results are represented as red squares. 
 We added the lithium abundances in unevolved stars from the literature  \citep{aoki2002, aoki2008, behara2010, bonifacio2012, bonifacio2015, caffau2012, caffau2013,  caffau2016, carollo2012, cohen2013,  frebel2007,   frebel2008, gonzales2008,  hansen2014, hansen2015, li2015, lucatello2003,     masseron2012, matsuno2017,  norris1997,  placco2016,  roederer2014,  sbordone2010,   sivarani2006,  spite2013}.

 Normal carbon stars are represented as black circles. Low and high carbon bands stars are shown as grey circles.   
The classification of CEMP stars follows the scheme  from \cite{bonifacio2018} who proposed :
 \begin{itemize}
 
 \item{} "carbon normal" : for [Fe/H] $\ge$ -4.0  [C/Fe] < 1.0 , for  [Fe/H] < -4.0 dex  A(C) < 5.5;
 
  \item{} low-carbon band CEMP stars: stars that do not fulfil the
carbon normal criterion and have A(C) $\le$ 7.6;
  
   \item{} high-carbon band CEMP stars: stars that do not fulfil the
carbon normal criterion and have A(C) > 7.6.
   
 \end{itemize} 

 Different symbols have been chosen for CEMP and non-CEMP  to show that the high-carbon band CEMP stars are preferentially Li-depleted whereas the lithium 	abundances in the low-carbon band are indistinguishable from that of the carbon normal stars. These measurements are consistent  with the hypothesis suggested by \citet{bonifacio2018} that the high carbon CEMP stars are the result of mass transfer from an AGB companion, also suggested by the works of \citet{koch2011} and \citet{monaco2012}.  
 
 We  evaluated the upper limit of the lithium  abundance  in all the stars of our sample. 
 4 of them reveal a lithium abundance lower than the Spite plateau value \citep{spite1982, bonifacio2007, sbordone2010}. These stars have metallicities below [Fe/H] =- 3.7 dex that place them in the region where the "meltdown" of the lithium plateau appears  \citep{sbordone2010, aoki2009, bonifacio2007}. The star SDSS~J081554.26$+$472947.5, identified as an EMP star by \citep{carbon2017, aguado2018},  has been studied in detail by \citet{hernandez2020} who found a metallicity of [Fe/H] =-5.49 dex while we found an upper limit of  -4.10 dex . It is interesting to note that, by adding more data in this metallicity range,  the  decrease of the lithium abundance is not constant at a given metallicity. In particular, two stars analysed by \cite{bonifacio2018} with a metallicity [Fe/H] $\simeq$ - 4.00 dex show a lithium abundance at the level of the Spite plateau.  From Fig. \ref{Fig_Lithium}, it seems that the Spite plateau appears as an upper limit for the lithium abundance in metal poor unevolved stars. 
 As the metallicity decreases below [Fe/H] =-3.0 dex, the dispersion of  the lithium abundance at a given metallicity seems to increase
 with decreasing metallicity. 

\subsection {Carbon}

\begin{figure*}
   \centering
  \includegraphics[width=12cm]{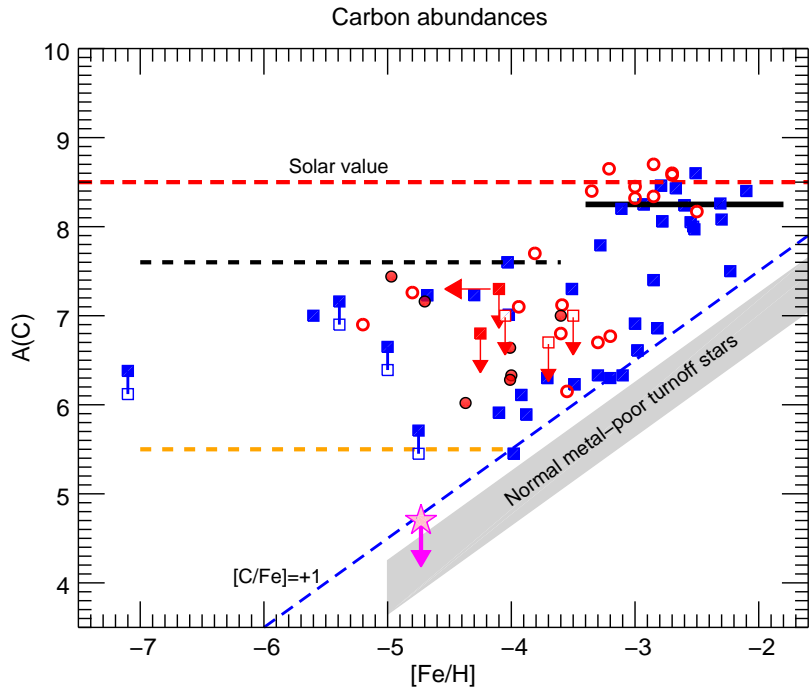}
      \caption{Carbon abundances A(C) of CEMP stars as a function of [Fe/H]. Large filled red squares represent our results. Down arrows indicate that the abundances are given as upper limits.Open red symbols represent the  stars for which we found with low [n-capture/Fe] upper limit abundances. Open and filled red circles are data published by our group. Other stars from the literature 
 \citep{sivarani2006, plezcohen2005, plez2005, frebel2005, frebel2006, thompson2008, aoki2008, behara2010, masseron2010, yong2013, cohen2013, li2015}      are represented as blue squares.    Open blue squares  are carbon abundances taking with  3D corrections from \citet{gallagher2016}. Black circles filled in red are stars from \citet{bonifacio2018}.  
       The pink symbol represents SDSS~J102915$+$17292, the normal carbon ultra metal poor star discovered by \citet{caffau2011a}.
      The other symbols are literature data. The black and yellow dashed lines  delimit the low-carbon band. Details can be found in \citet{bonifacio2018}.    }         
         \label{Fig_carbon}
   \end{figure*}

In Fig. \ref{Fig_carbon}, we plotted the abundance of carbon as a function of [Fe/H]. The results are represented as red squares with down arrows that indicate that the abundances are
upper limits.  Open red symbols represent the  stars for which we found with low [n-capture/Fe] upper limit abundances. 
We have also added literature results  \citep{aoki2008, behara2010,cohen2013, frebel2005, frebel2006,  li2015, masseron2010, sivarani2006, plezcohen2005, plez2005,  thompson2008,   yong2013} .  The  upper limits of the carbon abundance  we found in our five stars are compatible with them being moderately enhanced in C or C normal. As our measures are upper limits, the stars could be CEMP stars belonging tot the low  C band or just C-normal stars . 

\subsection {$\alpha$ and neutron-capture elements}

\begin{figure*}
   \centering
  \includegraphics[width=12cm]{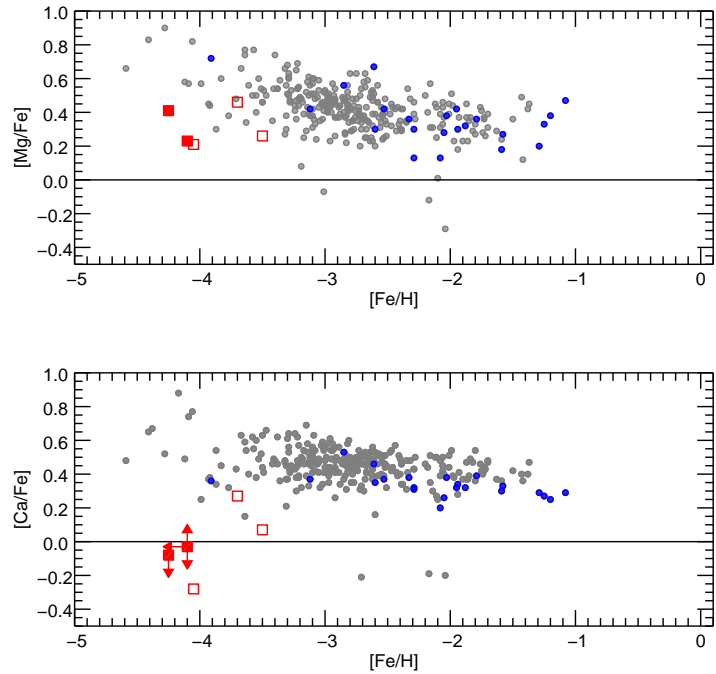}
      \caption{[Mg/Fe]  and [Ca/Fe] vs [Fe/H]. Red squares : this paper.  Open red squares represent the  stars for which we found with low [n-capture/Fe] upper limit abundances. Grey circles :  evolved stars from \citet{roederer2014}. Blue circles : main sequence stars from \citet{roederer2014}.  
              }
         \label{Fig_Alpha}
   \end{figure*}

\begin{figure*}
   \centering
  \includegraphics[width=12cm]{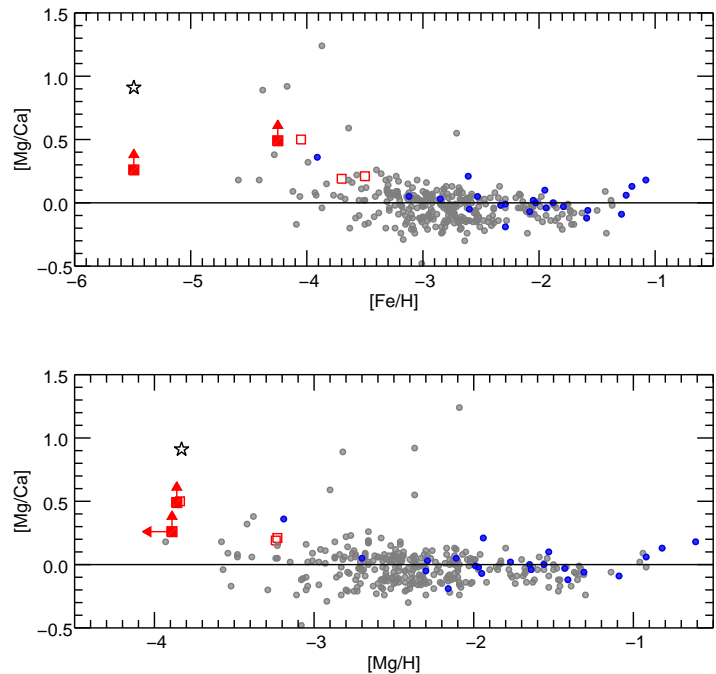}
      \caption{[Mg/Ca]  vs [Fe/H] and  [Mg/Ca]] vs [Mg/H]. Red squares : this paper.  Open red squares represent the  stars for which we found with low [n-capture/Fe] upper limit abundances. Grey circles :  evolved stars from \citet{roederer2014}. Blue circles : main sequence stars from \citet{roederer2014}. The black open symbol represents  SDSS~J081554.26$+$472947.5 \citep{hernandez2020} .  
              }
         \label{Fig_Subaru_MgCa}
   \end{figure*}

In Fig. \ref{Fig_Alpha}, we plotted the abundance ratios of [Mg/Fe] and [Ca/Fe] as a function of [Fe/H] for our sample of stars using the solar abundances from \citet{lodders09}.
We have added the literature data from \citet{roederer2014} that contain results from a rather large sample of evolved and main sequence stars. We could measure magnesium in 
the five stars.  Among them, SDSS~J081554.26$+$472947.5 seems to have a rather low [Mg/Fe] abundance ratio and a sub-solar [Ca/Fe] upper limit abundance. 
This is due to our  high upper limit of [Fe/H]. In fact,  by adopting the very low  [Fe/H] determined by \citet{hernandez2020}, we obtain  [Mg/Fe] = 1.62 dex in excellent agreement with the   [Mg/Fe] = 1.66 dex  abundance abundance they obtained.
The stellar parameters  we adopted for  SDSS~J081554.26$+$472947.5 is close to the one used by  \citet{hernandez2020} with a difference of 15K on the temperature, a difference
of 0.6 dex in log~g that has not a strong effect (typically 0.1 to 0.15 dex on the neutral Ca and Mg) on the determination of the abundance of neutral species and the same micro-turbulent velocity. 
Concerning calcium, we found three stars with sub-solar values, confirming the existence of  stars with low [$\alpha$/Fe] ratios as already suggested by \cite{bonifacio2018}.  However, we remind  that
 in very metal-poor stars, under the LTE hypothesis, the resonance line that we used to determine the abundance of calcium  leads to an underestimation of the calcium abundance \citep[and references therein]{spite2012}.  For turnoff stars, the amplitude of the effect is rather small.  \citet{spite2012} have computed a correction of the order of +0.1 dex in a turnoff star with [Fe/H] = -3.2 dex.
In Fig. \ref{Fig_Subaru_MgCa}, we plotted the [Mg/Ca] ratio respectively as a function [Fe/H] (upper panel) and as a function of [Mg/H] (lower panel). We have added the literature data from \citet{roederer2014} . 
The [Mg/Ca] ratio found in our sample of stars ranges around +0.2 to +0.5 dex. This ratio is slightly higher than the abundance ratio found in the [Fe/H] range -2 to -3 dex. At lower metallicities, the results from \citet{roederer2014} show that  the spread  in the [Mg/Ca] ratio  increases as [Fe/H] decreases  with values ranging from negative abundance ratios to highly enhanced  [Mg/Ca] ratios.  From our measurements,  the [Mg/Ca] ratio seems to increase as  [Fe/H]  decreases.  The result for SDSS~J081554.26$+$472947.5 found by  \citet{hernandez2020} seems to corroborate this point. The increase of the  [Mg/Ca] ratio is also visible when the ratio is plotted as a function of [Mg/H]. However, a high value of [Mg/Ca] at very low [Mg/H]  is not found in all the stars.  In particular, SDSS~J102915$+$17292 the normal carbon ultra metal poor star discovered by \citet{caffau2011a} has a solar [Mg/Ca] ratio. 

More observations at metallicities below [Fe/H] = -4 dex are necessary to conclude whether there is indeed an increase of the ratio [Mg/Ca] with decreasing metallicity or an 
increase of the spread of the [Mg/Ca] and a possible link to the C abundance found in the star.

\begin{figure*}
   \centering
  \includegraphics[width=12cm]{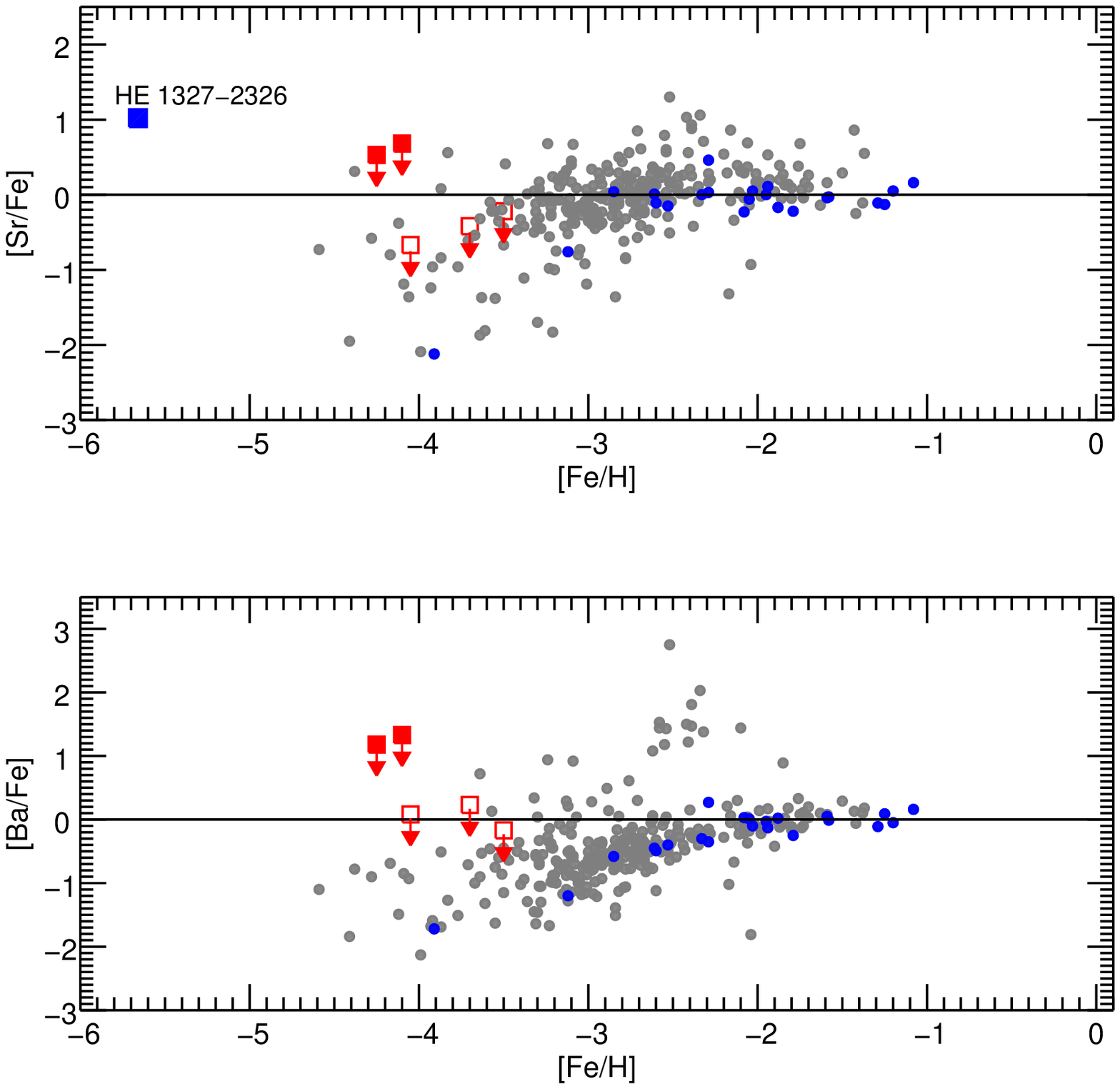}
      \caption{Sr/Fe]  and [Ba/Fe] vs [Fe/H]. Red squares : this paper. Open red squares represent the  stars for which we found with low [n-capture/Fe] upper limit abundances.   
       Grey circles :  evolved stars from \citet{roederer2014}. Blue circles : main sequence stars from \citet{roederer2014}. The blue rectangle represents HE1327-2326  a star with an exceptional high [Sr/Fe] ratio \citep{frebel2008, aoki2006}. 
              }
         \label{Fig_Ncapture}
   \end{figure*}

In Fig. \ref{Fig_Ncapture}, we plotted the abundance ratios and upper limits of [Sr/Fe] and [Ba/Fe] as a function of [Fe/H]. We have also added the results  from \citet{roederer2014} that contain results from a rather large sample of evolved and main sequence stars analysed in an homogeneous way.  Although we determine only upper limits, it is interesting to note that our stars can be divided into two groups, one with a high [n-capture/Fe] ratio, typically [Sr/Fe] $\simeq$ +0.5 to +0.7 dex and [Ba/Fe] around +1.2 dex   and the second one with solar [Sr/Fe]  $\simeq$ -0.4 to -0.7 dex and  solar [Ba/Fe] ratios. The low [n-capture/Fe] found  in  the group  of three stars is generally found in stars in the metallicity range around -2.5 dex.  The two stars  with very high [n-capture/Fe]  are similar to  the exceptional high [Sr/Fe] ratio  found in  the star HE1327$-$2326 \citep{frebel2008, aoki2006}. New observations with better S/N ratios would be very interesting to firmly determine the abundance ratios found in these stars.  

  %

\section{Conclusions}

In this article, we reported  the chemical analysis of five extremely metal poor candidates observed with the high dispersion spectrograph (HDS)
at the SUBARU telescope.  We discover a new UMP stars, SDSS~J105002.34$+$242109.7 with [Fe/H] =-4.25 dex. 
We could  determine the abundances of some elements (C, Mg, Ca, Sr and Ba) in the majority of these stars.  The five stars of the sample 
show abundance ratios which are typical of metal-poor stars in the metallicity range  -4.25 dex $\le$  [Fe/H] $\le$ -3.5 dex. These results show that the method 
developed by \citet{ludwig2008} to estimate the metallicity of unevolved stars from low resolution spectra is very efficient.  We could measure an upper limit of the  lithium abundance  in the five stars of our sample. 
The four most metal poor stars, with a metallicity ranging from -3. to -4.25 dex show lithium abundances below the Spite plateau. Some stars of our sample show a low [$\alpha$/Fe] content, a characteristic already found in previous studies \citep[an reference therein]{bonifacio2018}. For the star SDSS~J081554.26$+$472947.5 we obtain  a high  [Mg/Fe]= +1.62 dex  if we adopt the metallicity measured by \citet{hernandez2020}. 
  
\begin{acknowledgements}
 This work was supported by JSPS and CNRS under the Japan-France Research Cooperative Program (CNRS PRC No 1363), the JSPS Grants-in-Aid for Scientific Research (26400232, 26400237), and the RIKEN iTHEMS Project.
P.F. acknowledges support by the Conseil Scientifique de l'Observatoire de Paris (AFE programme).  
We gratefully acknowledge support from the French National Research Agency (ANR) funded project ‘Pristine’ (ANR-18-CE31-0017).
\end{acknowledgements}

%
%

\end{document}